\documentclass[conference]{IEEEtran}
\ifCLASSINFOpdf
   \usepackage[pdftex]{graphicx}
   \DeclareGraphicsExtensions{.pdf,.jpeg,.png}
\else
\fi

\usepackage{algorithm}

\usepackage{algpseudocode}
\usepackage{siunitx}
\usepackage{subfig}
\usepackage{cite}
\usepackage[final]{changes}

\newcommand{\enesays}[1]{\color{red} #1 \color{black}}
\renewcommand{\enesays}[1]{}

\definechangesauthor[name={Dominik}, color=red]{dominik}
\definechangesauthor[name={Suat}, color=blue]{suat}
\definechangesauthor[name={Mumin}, color=purple]{mumin}
\definechangesauthor[name={Enes}, color=purple]{enes}
\usepackage{atbegshi,picture}
\usepackage{lipsum}

\AtBeginShipout{\AtBeginShipoutUpperLeft{%
  \put(\dimexpr\paperwidth-1cm\relax,-1.5cm){\makebox[0pt][r]{\framebox{Part of this paper will be presented in IEEE ICBC, 2020, Toronto}}}%
}}

\begin{document}
%

\title{A Cost-efficient IoT Forensics Framework with Blockchain}

\author{\IEEEauthorblockN{Suat Mercan, Mumin Cebe, Ege Tekiner, Kemal Akkaya, Melissa Chang and Selcuk Uluagac} 
\IEEEauthorblockA{Dept. of Elec. and Comp. Engineering\\
Florida International University\\
Miami, FL 33174\\ Email: \{smercan,mcebe,etekiner,kakkaya,mechang,suluagac\}@fiu.edu}
}

\maketitle

\begin{abstract}

IoT devices have been adopted widely in the last decade which enabled collection of various data from different environments. The collected data is crucial in certain applications where IoT devices generate data for critical infrastructure or systems whose failure may result in catastrophic results. Specifically, for such critical applications, data storage poses challenges since the data may be compromised during the storage and the integrity might be violated without being noticed. In such cases, integrity and data provenance are required in order to be able to detect the source of any incident and prove it in legal cases if there is a dispute with the involved parties. To address these issues, blockchain provides excellent opportunities since it can protect the integrity of the data thanks to its distributed structure. However, it comes with certain costs as storing huge amount of data in a public blockchain will come with significant transaction fees. In this paper, we propose a highly cost effective and reliable digital forensics framework by exploiting multiple inexpensive blockchain networks as a temporary storage before the data is committed to Ethereum. To reduce Ethereum costs, we utilize Merkle trees which hierarchically stores hashes of the collected event data from IoT devices. We evaluated the approach on popular blockchains such as EOS, Stellar, and Ethereum by presenting a cost and security analysis. The results indicate that we can achieve significant cost savings without compromising the integrity of the data.

\end{abstract}

%
\maketitle

\IEEEpeerreviewmaketitle

\section{Introduction}

The advancement in communication technologies, sensing items, and
affordable computing devices has led us to the age of internet of things (IoT) which enables collecting diverse ambient data and communicate it to remote locations \cite{iotdef}. IoT is becoming the de facto technology in many domains including transportation, energy, healthcare, agriculture, hospitality, etc. \cite{iotapp}. In these applications, the data collected from various IoT devices are used to conduct extensive analytics to make informed decisions and take actions. In some scenarios, however, the data is very crucial to run critical infrastructure (i.e., power systems, transportation) and understand failures when they occur. In particular, if there are failures due to human errors or deliberate attacks, it is utmost important to be able to detect the cause of these failures and hold involved parties responsible. Therefore, the secure transmission and storage of IoT data is critical for such purposes \cite{aman}. This necessitates mechanisms to be able to store IoT data for digital forensics investigation purposes. As the data needs to be presented as evidence in case of disputes, there is a need for trustworthy storage which cannot be deleted or modified. Emerging Blockchain technology can be an excellent fit for such scenarios since it can provide authenticity verification, data provenance, and data integrity \cite{li}. It comes with a distributed ledger technology which can run consensus algorithm among peers to enable transactions in trustless environments. This eliminates the need for a central authority and thus provide a distributed trust. Indeed, with such features Blockchain technology \cite{bitcoin}  has opened doors to many novel applications in various domains\cite{blockchainapp}. Among these, forensic investigations, healthcare, insurance business etc. \cite{li} are of interest since there is a need to prove that the stored data has not been tampered with after it was saved.


As a more specific example, let us consider rental businesses and insurance industry which can rent cars or boats.  
When a person rents a vehicle or any other asset, a dispute might occur among the stake-holders in case of an accident, failure or illegal usage. The renter must operate the vehicle by complying with the regulations, and an insurance company may want to ensure that they are covering only what they are responsible for. The insurance companies have to deal with fraudulent claims valued at millions of dollars every week \cite{insurance}. The company must validate if the preconditions of the policy are met. In order to establish a ground where everyone is held accountable fairly, the data generated by sensors/IoT devices must be recorded timely, stored transparently and securely. We argue that blockchain technology can address this issue~\cite{blockchaininsurance}.

Specifically, an ideal solution would be creating a permissioned blockchain (i.e. a private blockchain network) which allows only certain entities to join the network where some untrusted parties can exchange information. Stakeholder such as users, rental company, renter, and insurance company can become part of this private blockchain network. However, this approach brings alot of overhead in terms of managing the private ledger. Thus, the stakeholders would not be not cooperative in this respect. In addition, the security of a private blockchain depends on the number of users and small ones could be risky in terms of consensus. Thus, it makes more sense to use a public blockchain to eliminate the management overhead and increase trust.



However, in the case of public blockchain, there is the challenge of costs with transactions. This is particularly the case with popular blockchain networks such as Ethereum or Bitcoin. If huge amounts of data is to be written to these public ledgers, this may annually cost a lot of money even for Ethereum which is much affordable compared to Bitcoin \cite{price}. In addition, the solution would not be scalable as the number of IoT devices writing to blockchain increases dramatically. While it is possible to use other less costly ledger platforms instead of Ethereum, their reliability will be much less since these ledgers may not have enough nodes and thus 51\% attack may be performed with less effort. Therefore, there is a need for cost-effective mechanisms to store IoT data in public blockchains. 



In this paper, we design and evaluate a forensics framework for IoT data integrity verification by proposing a multi-chain approach where we utilize multiple relatively affordable  blockchain networks such as EOS \cite{eos} and Stellar \cite{stellar} (compared to Ethereum and Bitcoin) for temporarily storing the hash of the IoT data before they are permanently stored to Ethereum. To reduce the hash sizes further, we propose to utilize Merkle trees that can represent a number of hashes in a single hash value stored in a tree-like structure. The overall idea is creating a secure platform with the combination of several blockchains which makes it more powerful than the sum of each individual.

Specifically, as the data collected from an IoT device during predefined events, it is transferred to company database. Then, the hash of that data is stored in EOS  and Stellar blockchain ledgers which serves as the first security level in our proposed framework. At the end of each day, the IoT device retrieves the block information from these two platforms and inserts them in a Merkle tree whose roots are written to Ethereum as a second level security platform. The advantages of this approach are twofold: First, it is cost-efficient because we are saving the incident information mostly in EOS and Stellar, which are much cheaper and only daily summary of all transactions is written to Ethereum. 
Second, the proposed framework has higher security and resiliency against \%51 consensus attack \cite{51attack}. The attacker must hack both blockchain networks in the first level within the same day before the summary is written to Ethereum, or s/he has to change data both in Ethereum and one of the blockchains in the first level. Their  consensus algorithms make it even more difficult to launch an attack. Note that the number of blockchains in the first level can be increased to further strengthen the security of the system. We showed through a cost analysis that the proposed framework can reduce the costs by more than 10 folds of magnitude. 

This paper is organized as follows: In the next section, we summarize the related work in the literature. Section III provides some background on the used concepts. Section IV presents the system and threat model while Section V details the proposed scheme. In Section VI, we present a cost and security analysis of the proposed mechanism. Section VII concludes the paper.

\section{Related Work} 

Various ideas have been proposed related to blockchain utilization for data integrity verification.  The works done by \cite{olufowobi,polyzos} focus on generic blockchain-based data provenance infrastructure for IoT generated data. They first identify the security and trust related challenges, and then disscuss how blockchain can be included to overcome these issues. Authors in \cite{block4forensic} suggest a framework for car accident scenarios which supposed to save data in blockchain when an accident happens. They use a simplified public key infrastructure tailored for vehicular networks to preserve the privacy. The data saved in blockchain is used to solve any dispute among the insurer, owner and manufacturer. Gipp et al. \cite{gipp} implemented a similar approach on smartphones which is used as dashboard camera in cars. Once smartphone detects an accident via accelerometer sensor, it starts recording the scene and calculates the hash at the end to be written to the public blockchain. In order to keep the cost to minimum, it stores the aggregation of the hashes. In order to prove that the video stored on the phone has not been changed, the user can provide the original video with the hash. Block-DEF\cite{blockdef} is proposing a secure digital evidence framework using blockchain. The idea is to store evidence and evidence information separately. In order to avoid data bloat, they are proposing a lightweight blockchain design. They claim that it is a scalable framework to keep the evidence safe and tamper-proof. ProvChain\cite{provechain} try to provide data assurance for the collected data through IoT sensors. They calculate the hash of the data and store it on the blockchain network instead of the whole data. Our work differs from previous works in two ways: First, in addition to hash-based storage we utilize Merkle trees to further save space. Secondly, and more important, we use multiple low-cost blockchain networks collaboratively to increase the reliability/security while keeping the cost lower.

\section{Background}
\label{sec:background}
In this section, we provide some background on the concepts used in our approach. 

\subsection{Blockchain}
\begin{figure}[h]
\begin{centering}

\includegraphics[scale=0.47]{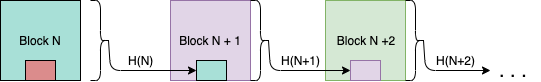}
\caption{Blockchain}
\vspace{-2mm}
\label{fig:blockchain}
\end{centering}
\end{figure}

Blockchain is a list of records called blocks, first proposed by Satoshi for Bitcoin \cite{bitcoin} which became popular quickly in the world. The aim is providing decentralized trust. 
Blockchain technology is a combination of various technologies such as cryptographic hash algorithms, peer-to-peer (P2P) distributed network data sharing, digital signatures, and Proof of Work (PoW) consensus protocol. Cryptographic hash algorithms provide us data integrity and Blockchain uses this feature to bind chains together by their hash values as hown in Fig. \ref{fig:blockchain}. P2P distributed network model provides decentralized communication among nodes. With digital signatures, nodes can manage their assets and prove their possession without relying on a central authority. PoW consensus protocol is the main innovation of Blockchain technology which guarantees randomness and decentralized reward election. 

Note that blockchain can be used to implement tamper-resistant data storage. Once a data is deployed into the blockchain, it is almost impossible to change this data in large-size blockchain networks such as Bitcoin and Ethereum. These networks have thousands of nodes (115,000 for Ethereum and 100,000 for Bitcoin) for storing data into their own ledgers. If any malicious person wants to change the data in these ledgers, s/he must change at least 51\% of the nodes. This attack is referred to as \textit{51\% attack} in the rest of the paper. This feature provides a high level of data integrity. 

\subsection{Ethereum}
\begin{figure}[!ht]
\begin{centering}
\fbox{
\includegraphics[scale=0.4]{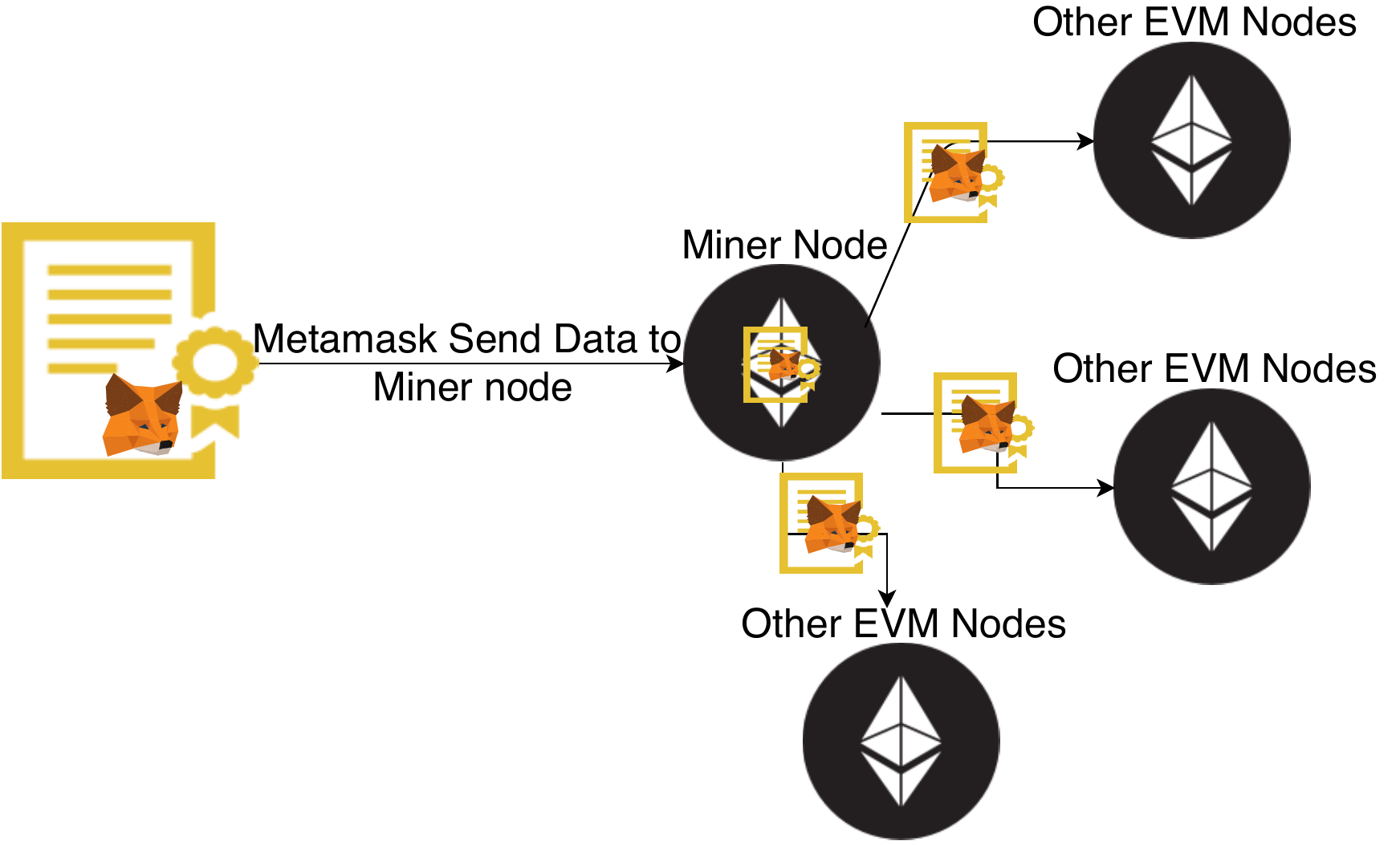}}
\caption{Ethereum Smart Contract Distribution}
\vspace{-0mm}
\label{fig:ethereum}
\end{centering}
\end{figure}

Ethereum\cite{ethereum} is the most widely used and trustable Smart Contract oriented Blockchain network in the world. It is a public, permissionless blockchain which means that anyone can access the information on Ethereum and initiate transactions on their own. It was built as a platform for people to develop decentralized applications easily. Instead of storing monetary transactions in the blocks, one can store smart contracts, or code snippets, in the blocks as seen in Fig. \ref{fig:ethereum}. Ethereum uses solidity programming language for creating contract which is compiled by Ethereum virtual machine. Every contract has a \textit{gas} fee that is calculated based on the contract’s memory space and total workloads. It becomes more expensive when the data size gets bigger. Ethereum currently uses PoW consensus algorithm like Bitcoin but Ethereum's block frequency is between 10-20 seconds since its hash puzzles are much easier to solve. Thus, Ethereum generates blocks faster and has higher throughput. 

\subsection{Stellar}
Stellar\cite{stellar} was the first smart contract oriented blockchain network which aims to provide a scalable payment gateway for users. Stellar is very scalable as its block/contract mining time is around 3-5 seconds. It can confirm thousands of transactions per second. Stellar uses gossip network-based voting algorithm, named Stellar Consensus Algorithm, for consensus in the blockchain network. The development team also provides a wide range of API/SDK to make better development in Stellar blockchain. Finally, multi-signature enabled smart contracts are beneficial for multi-user applications. Lumen is used as a cryptocurrency in Stellar blockchain network. Due to Lumen's low currency exchange rate, the contract creation and deployment fees are so low making Stellar an affordable blockhain option.  

\subsection{EOS}
EOS\cite{eos} is a well-known and efficient Blockchain Network. Its name comes from Ethereum Operating System. EOS uses delegated proof of stake as a consensus protocol that provides high efficiency and low energy consumption. Deploying smart contract to EOS network is easy and free but the contract creator should hold some amount of EOS, CPU, and RAM to use EOS bandwidth efficiently. One central EOS full node is enough for multiple wallets.

\subsection{Smart Contract}
Smart contracts are pieces of code that are executed by virtual machines which are run in all full blockchain nodes in the network. These Virtual machines are generally compilers which collaborate with the public ledger of the node. When the contract creator deploys a smart contract, the metadata of this smart contract is broadcast to all nodes in the network and becomes un-erasable. The way these smart contracts can be used varies based on the platform. Smart contracts can be utilized to implement various use cases by eliminating 
third parties. For instance, people can exchange any asset such as a vehicle without involving the government authority since they can prove the ownership of this vehicle by using the records on the distributed ledger. Other use cases include rule-based transactions which are achieved using some \textit{if and else statements} in these contracts. 
Smart contract concept has great to potential to ease some daily operations though governments do not have any regulation yet for smart contract uses.

\subsection{Merkle Tree}
Merkle tree \cite{merkle} is a fundamental data structure that allows effective and reliable verification of content in a huge collection of data. This structure serves to check the consistency and content of the data. Basically, a Merkle tree compiles all the data in a tree by producing a digital fingerprint of the entire set, thereby allowing any actor to verify whether or not a specific node is included in the tree.

\begin{figure}[!ht]
\begin{centering}
\fbox{
\includegraphics[scale=0.45]{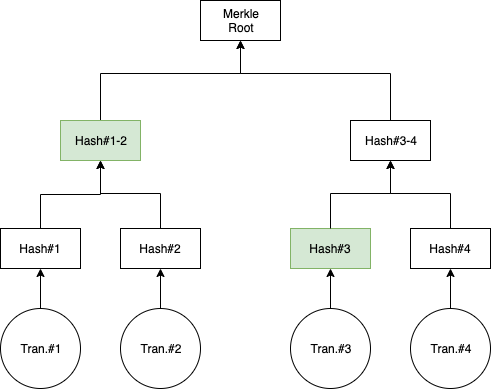}}
\caption{Merkle Tree}
\vspace{-2mm}
\label{fig:merkle1}
\end{centering}
\end{figure}

Merkle trees are formed by repeatedly hashing nodes until there is only one hash left, which is called Merkle Root as seen Fig~\ref{fig:merkle1}. The hashing is conducted from the bottom up, starting from hashes of individual data points. Specifically, each leaf node in the tree is a different hash of data point, and each non-leaf node is a hash of its two separate roots of its sub-trees.  Merkle tree is actually a perfect binary tree structure which requires an even number of leaf nodes. Thus, when the number of individual data points is odd, simply the last data point is duplicated to produce an even number of leaf nodes.  

One of the essential differences of a Merkle tree from a basic hash-list is that branches can be fetched independently. As a result, the integrity of each branch can be verified independently. This is beneficial because records can be split up into small data chunks such that only a small piece needs to be fetched to verify the integrity of any leaf node. This process is called simplified verification (SV) and proves that particular transactions are included in a Merkle tree without downloading the entire tree.

As an example consider SV of Tran.\#4 in the Merkle tree given in Fig. \ref{fig:merkle1}. If a verifier wants to check whether Tran.\#4 is included in the Merkle tree, it just needs to fetch the shaded hash values in the tree (i.e., Hash\#1-2 and Hash\#3). Using these hash values, s/he can re-compute his/her Merkle root and compare with the the given Merkle root. Specifically, the process works as follows: 
\begin{enumerate}
    \item The verifier aggregates Hash\#3 (given) and hash(Tran\#4) which is available to derive Hash\#4. 
    \item The verifier aggregates the given Merkle path node, Hash\#1-2 and Hash\#4 to derive Hash\#3-4. 
    \item The verifier aggregates Hash\#3-4 with the given Hash\#1-2 to derive the Merkle root. 
    \item The obtained Merkle root is compared with the given Merkle root. If they match, the verification is complete. 
\end{enumerate}


Consequently, the Merkle tree significantly reduces the amount of data in a verification process. It provides a unique mechanism to provide a validation process without requiring the whole data itself. For instance, to prove that a continuously growing transaction log is complete and intact, the Merkle tree provides a lightweight verification method which includes all previous transactions. The verification ensures no previous transactions have been altered, and the tree has never been branched.  Because of this unique verification, Merkle tree benefits both provers and verifiers. A prover can compute hashes progressively, as it collects new transactions. A verifier can verify a transaction individually by checking individual hashes of other branches of the tree.

\section{System and Threat Model}

\begin{figure}[!ht]
\begin{centering}
\fbox{
\includegraphics[scale=0.4]{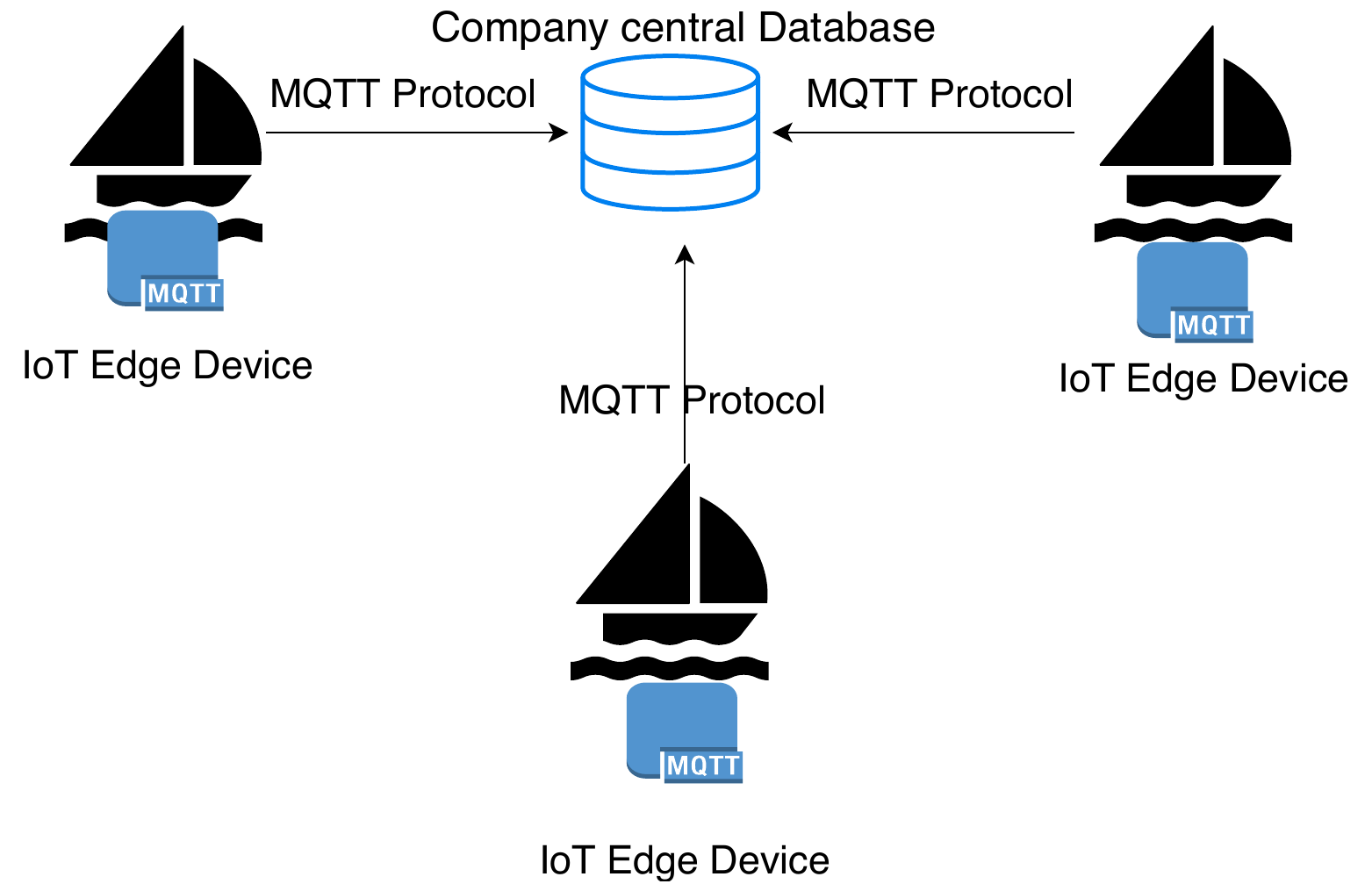}}
\caption{System Model}
\vspace{-0mm}
\label{fig:initialsystem}
\end{centering}
\end{figure}

In this paper, we consider a boat rental application, which is very common in South Florida. Basically, boats are rented by a boat rental company and their data are collected via on-board sensors. Each boat is equipped with an on-board IoT edge device that can communicate with various sensors within the boat using CAN bus protocol. All the sensor data may not be equally significant for the rental company though. Thus, it is not required to write all of them to blockchain or even to database. The data is filtered out based on significance or certain events. For instance, the renters are allowed to drive within a specific zone for which the insurance is valid. So, it may not be necessary to transmit the geolocation constantly, but if an accident happens or when the boat goes outside of designated area, the data becomes important. When the system decides that a data is important, then it is transmitted to a remote company database by the edge device through the widely used MQTT protocol \cite{mqtt} and 4G/LTE communication. A sample system model is shown in Fig. \ref{fig:initialsystem}.  

The security of the proposed forensic framework depends on the secure implementation of proposed multi-chain system. Therefore, we consider the following threats to the security of the proposed approach and identified the relevant security goals. Note that in our attack model, we assume that IoT edge device is tamper-proof through Hardware Security Modules (HSMs) that provide device-level controls to protect deployed IoT devices. Therefore, IoT edge device infiltration is out of scope.  

\begin{figure*}[!ht]
\begin{centering}
\fbox{
\includegraphics[scale=0.50]{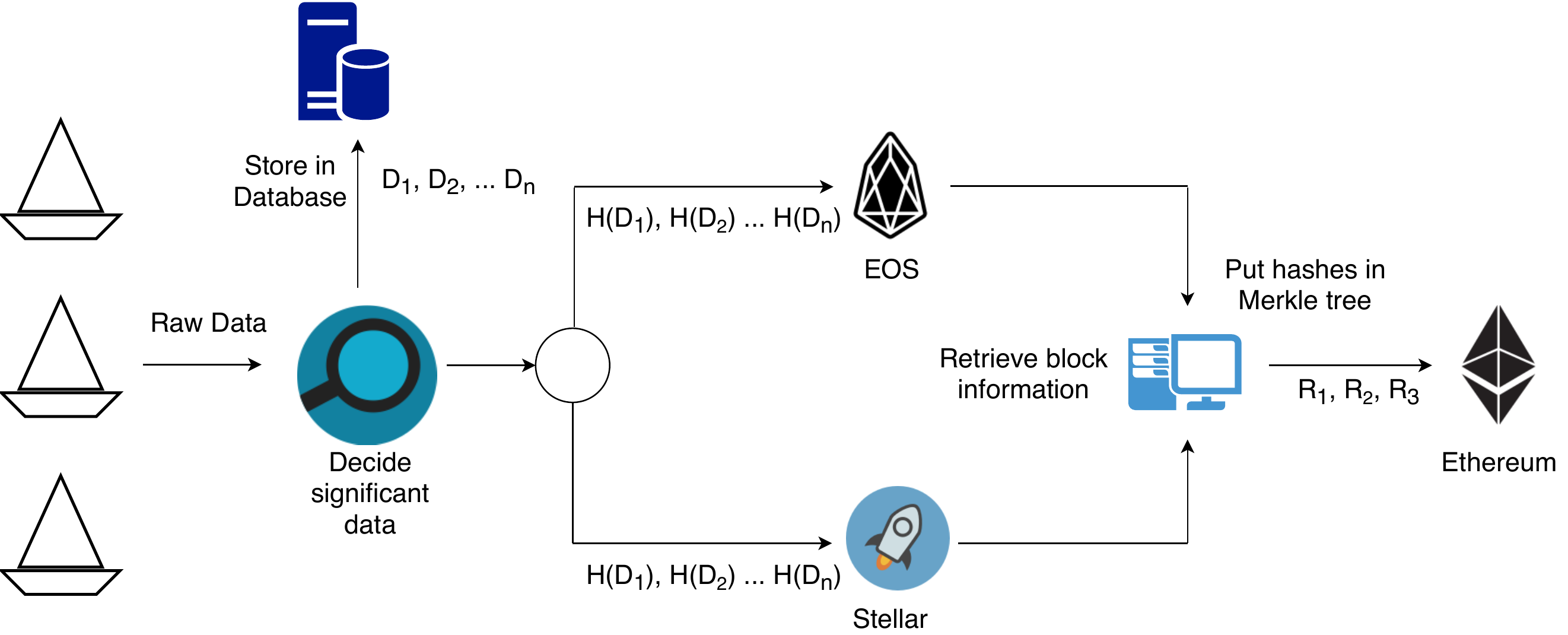}}
\caption{Proposed System Design}
\vspace{-6mm}
\label{fig:diagram}
\end{centering}
\end{figure*}

\textbf{Threat 1:} In this scenario, the attacker disguises itself as an IoT edge device for pushing false  IoT data into our multi-chain system by inferring the private keys that are used to sign the data.

\textbf{Threat 2:} In mining-based consensus protocols, all the data is kept in the memory pools (\textit{mempools}) and miners choose transactions from this memory area. In other words, the mempool is basically the node’s waiting room for all the unconfirmed transactions. Each blockchain node has a different mempool size for stocking the transactions. Thus, each node has a different version of the pending transactions. This creates a variety of pending transactions on different nodes in a distributed manner. If the size of unconfirmed transactions becomes too large to fit in memory, the miners independently remove some of the pending transactions from their mempool. To do so, miners generally remove the transactions which pays less transaction fees to boost their earning from mining. In this scenario, the attacker attacks the mempool and get the transactions which have the hash values to be removed or delayed.


\textbf{Threat 3:} In this attack scenario, the attacker attacks the IoT device communication layer and performs a man-in-the-middle (MitM) attack for altering the  transactions.

\textbf{Threat 4:} In this attack scenario, the attacker can counterfeit data in the Ethereum Blockchain.

\section{Proposed Approach}
\label{sec:pre}

\subsection{Motivations}
In our case, we seek an efficient solution for a  boat rental company which aims to store sensor data collected from its boats in a such a way that it can prove the integrity of data in future retrievals. Note that a secure integrity mechanism would not only reduce their insurance rates but also help them in quickly resolving potential disputes with its customers. 

There may be different options to tackle this issue by leveraging the blockchain technology. As mentioned, blockchain is a promising environment to verify and prove the integrity of prerecorded data. It has great potential specifically for insurance industry since registering important data will be beneficial in resolving disputes among stake-holders. Therefore, the most obvious option would be to construct a permissioned blockchain among stake-holders. This type of solution is applicable when multiple untrusting parties want to share information. For instance, raw material provider, manufacturer, transporter, seller in a supply-chain link can create a consortium for data provenance and integrity. IBM's Hyperledger \cite{hyperledger} is designed for this type of business cases. For our specific case, insurance companies are typically not cooperative due to management costs which rules out this option. However, since the rental company still wants to store data in an immutable way, utilizing a public a blockchain could be an option. 

Therefore, the second solution could be to write the data directly to Ethereum network which is a highly secure 
blockchain platform. It will be required to have stake worth billions of dollar to make a 51\% attack. However, writing every single transaction on Ethereum will be highly costly considering the number of transactions in IoT cases. Ethereum might be feasible for some other cases such as asset transfer utilizing smart contract. For instance, when the ownership of car is transferred, the money transfer will be completed. However, in cases where we need frequent transactions, it is not a very cost-efficient method. 

Another option would be that data is saved in the database and the calculated hash of stored data is written to Ethereum periodically (i.e., once a day). This will reduce the cost significantly when compared to previous approach and ensure the data integrity after it is written to blockchain. While this reduces costs, it does not guarantee the data immutability for the duration of data residing in the data center database. So, this approach has issues with the security while the cost is lower.

Therefore, we opt for a more cost-efficient approach that will rely on multiple blockchain networks as detailed next.

\subsection{Proposed Multi-chain Framework}
One of the biggest challenges with traditional forensics mechanisms is the need to maintain an additional trusted authority to ensure the integrity of the data. Regardless of being encrypted or not, if the trusted authority is compromised, it provides an intruder with an origin to play with the integrity of the data. In addition, a single trusted authority alone cannot stand for an insider attack when it becomes a target of interest.  

\begin{figure*}[!ht]
\begin{centering}
\fbox{
\includegraphics[scale=0.65]{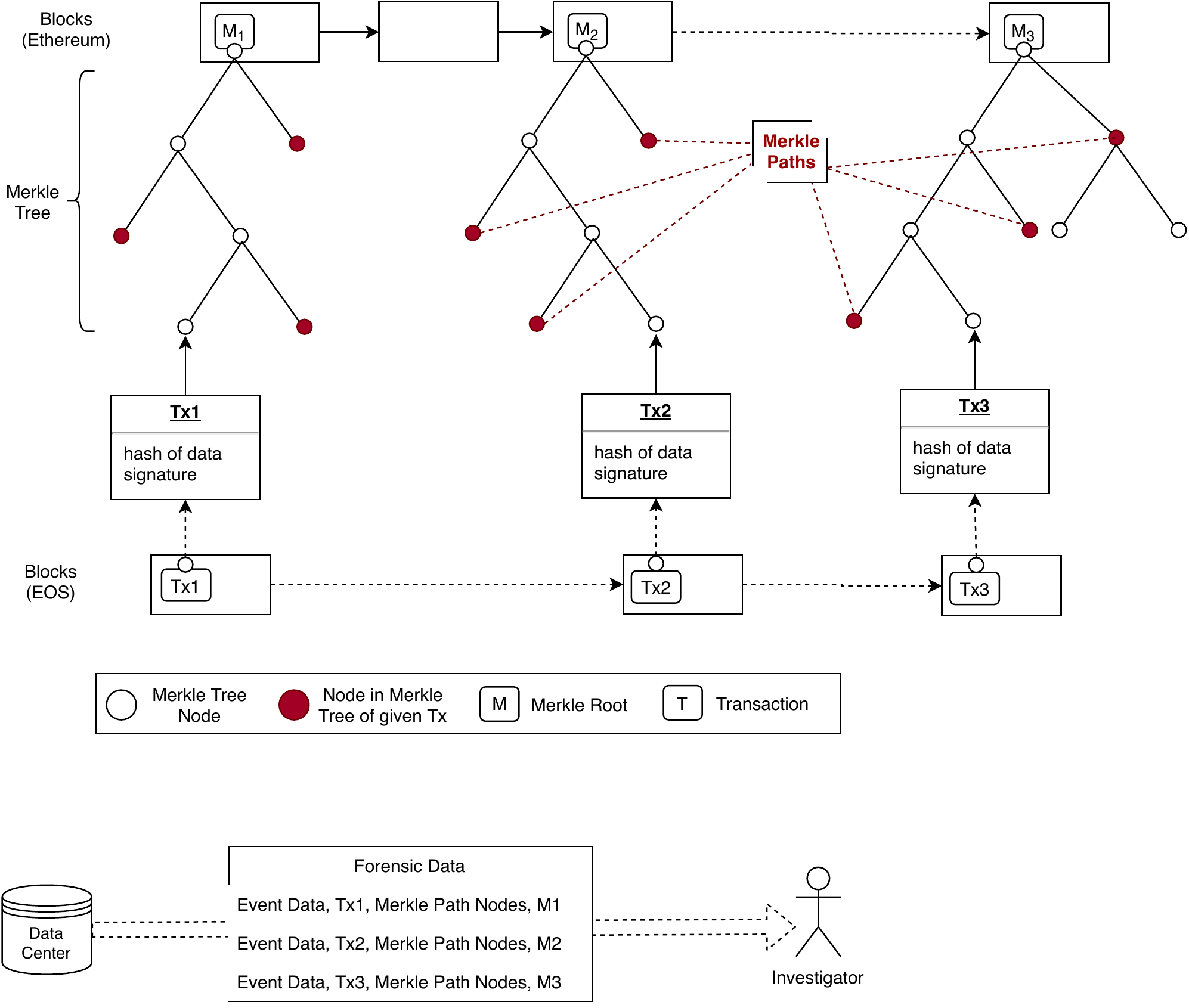}}
\caption{Integrity Verification Process}
\vspace{-2mm}
\label{fig:merkle}
\end{centering}
\end{figure*}

Our proposed framework, first, alleviates the trust issue and establish a trustless setup by utilizing the blockchain technology. However, storing data on public blockchain is both expensive and has lots of privacy concerns. As mentioned, even though the boat rental company can maintain only the hash of the data in the blockchain to compare it with the actual data on its database during forensic investigations, keeping just the hash of IoT event data on a secure and public blockchain still costs a lot of money in the long run as will be shown in the Experiment section.

Therefore, we resort to affordable alternatives for public blockchains. Although Bitcoin and Ethereum are most popular and sustainable blockchain platforms, there are many others such as Stellar and EOS, which have survived for years. While they may not be as reliable as Bitcoin and Ethereum because of the limited number of users and popularity, the cost of using these platforms are significantly lower. Since using only one of these platforms may not be secure, we propose a multi-factor integrity (MFI) system that utilizes multiple low-cost blockchain platforms together with Ethereum. The idea is similar to using a back-up system against system failures. We aim to increase the resiliency of data in case one of the platforms will discontinue or hacked. MFI makes it more difficult for a malicious actor to alter the IoT data in a stealthy way, which is stored in the company's database. If one blockchain is compromised or broken, a malicious actor still has at least one more obstacle to breach the integrity of the data. Note that these platforms are all smart contract oriented to enable easy communications among each other.

To reduce data size to be written to public blockchains, we employ hash functions along with Merkle tree. Our approach for reducing costs is as follows: 
\begin{enumerate}
\item In the first step, the IoT edge device from a boat submits the hash of IoT data to the first level of the multi-chain system. As mentioned, only interesting data is picked based on some predefined events or pre-conditions. The hash of this data is written to both Stellar and EOS during the day as long as there is inteersting  data. 

\item At the end of each day, a synchronization process starts and the data center of the rental company fetches the confirmed transactions which were submitted to the first level blockchains Stellar and EOS. The data center then builds a Merkle tree from confirmed transactions and computes the Merkle root for each. 

\item The Merkle root computed in the previous step represents another integrity factor. Thus, it is submitted to a more secure and reliable blockchain, Ethereum, and a copy of it is stored in local database to be used during forensic investigations. Note that Ethereum is used only for the hash of all hashes during a day to save transaction costs. 
\end{enumerate}

The proposed overall architecture is shown in Figure \ref{fig:diagram}.

\subsection{Integrity Verification Process}

When there is an incident that creates dispute on who is responsible, the proposed framework will be used to find out and prove what happened. Basically, a an insurance company working on a claim or a law enforcement officer working on an accident scene needs to ensure the integrity of the available data. Once the integrity of the data is ensured, the faulty party can be determined undeniably.

To do so, considering our framework, the investigator/officer first accesses the related forensic data that is stored in the data center. S/he will then need to collect the submitted transactions that contain the hashed of the data to the first level blockchains, related Merkle root values and Merkle paths of those transactions. 

The details of the process is shown in Figure \ref{fig:merkle}. In this figure, the data center contains a row for each event data which includes: 1) the original IoT data; 2) The transaction created with the hash of this data; 3) The Merkle path nodes for this data; and 4) The Merkle root.


Let us consider Transaction  \#1 (Tx1). This transaction contains the hash of an event data that is already stored in the data center. The investigator/officer can trust the event data if it exists in the first level blockchain. Basically, the hash of the event data is computed and this hash is compared with the value in Tx1 in EOS and/or Stellar. 

Then, the investigator/officer, can of course, choose to validate the input transactions again on Ethereum. In order to validate Tx1 and ensure that it exists within the Merkle path, s/he needs to check whether the provided Merkle root which contains the Tx1 and given Merkel root M1 is equal to the value stored in Ethereum. To do so, s/he simply needs to compute the hash of Tx1 that acts as a node in the Merkle tree provided by the data center. With the given nodes in its Merkle path, the investigator/officer can simply and very quickly calculate M1 (i.e., the Merkle root of which Tx1 belongs to) by series of hash operations with SV method as described in Section III. If the calculated Merkle root by investigator is equal to the provided M1, s/he ensures that the path in the Merkle tree is correct and Merkle root M1 contains Tx1. 

If the computed Merkle root and the value which is saved in the Ethereum matches, the investigator/officer knows with certainty that the data center has given him/her a valid/tamper-proof IoT hash data. S/he also knows that the existence of the transaction in the blockchain has been validated by different multi-chain miners and that there is an extensive PoW/computation time ensuring the integrity of the hash data in the multi-chain system. The overall process of verification is shown in Figure \ref{fig:merkle}.

\section{Performance Evaluation}

In this section, we evaluate the proposed framework in terms of its associated costs. 

\subsection{Experiment Setup}
To perform transactions in each of the mentioned blockchain platforms, we created their respective nodes. First, we setup EOS central node to which nodes are supposed to connect through the EOS chain plugin API. Every node (boats in our case) has its wallet and uses this wallet to connect to main EOS Network. When there is any interesting event, IoT edge device pushes the hash of this value into EOS Blockchain via Central EOS Node. In this setup, it is possible for each wallet to track all transactions easily via history\_api\_plugin. Similarly, we installed Stellar wallets to connect to its network using theirs APIs. 


We use events waiting for API recalls to trigger smart contracts which becomes ready to be deployed to Ethereum after checking validity. Event is an interface between wallet, API, and smart contracts. Javascript API connects to Web3 interface of Ethereum client that hosts smart contract, and triggers the hash deployment event. Triggered function imports the Merkle root data that we constructed, and creates a transaction for Metamask wallet which is the most widely used wallet by Ethereum developers. Metamask wallet broadcasts metadata of the contract to all main Ethereum network via peers, and wait for one miner to put the smart contract into a block.  


\begin{table}[h]
\renewcommand{\arraystretch}{1.5}
\caption{Transaction Cost}
\label{table1}
\centering 
\begin{tabular}{|c|c|c|c|c|}
\hline
 \textbf{Blockchain} & \textbf{Unit Cost} &  \textbf{Cost in \$}  &  \textbf{Time}\\
\hline
\hline
 EOS & 100 EOS (once) & \$0.00063 & $<$ 1 min  \\
\hline
\hline
 Stellar & 0,001 Lumen & \$0.000054 & $<$ 1 min \\
\hline
\hline
 Ethereum (contr) & 0.000131 ETH & \$0,019 & 12 min  \\
\hline
\hline
 Ethereum  & 0.000025 ETH & \$0,0036 & 10 min \\
\hline
\end{tabular}
\end{table}

\subsection{Benchmarks}
We compared our approach with two other approaches as described below: 

\begin{itemize}
\item \textit{Ethereum  with  new  Contract}: This  approach  creates  a  new Ethereum contract  for  each  hash and insert the hash in this contract.  Note that creating  a  new  contract  for  each  piece  of  data  is costly  but  it  is  simple  and  most  secure  way  to  store  data  in  the  Ethereum  Blockchain.   

\item \textit{Function Call from Ethereum Contract:} This approach deploys an Ethereum contract by including a function, and thus each time this function is called to save the hash instead of creating a new contract. Making a function call is a cheaper process than a new contract deployment since the contract creator pays only the contract creation fee once. However, when  one  smart contract  is  deployed  and  its  function  is  called  to  save  a  new hash  value,  it  becomes  less  secure.  The  attacker  can  directly attack  a particular  contract  instead  of  hundreds  of  them.  

\end{itemize}
\subsection{Cost Analysis}
We assessed the cost associated with the proposed framework by comparing it with the benchmarks mentioned. Before doing the complete cost analysis, we measured and provided the unit transaction costs associated with each blockchain platform for a function call to save a hash value along with the transaction verification time in Table \ref{table1}. We observe that Ethereum unit price, even deployed with the minimum gas fee, is much higher than others. It should also be noted that EOS provides free contract deployment but it requires to have 100 EOS in the node. The other note is about the validatin times. EOS and Stellar are much faster for real-time transactions. Ethereum on the other hand is slow but since it is used at the end of the day on already stored transactions, this would not be of concern. 


\begin{table}[!ht]
\renewcommand{\arraystretch}{1.5}
\caption{Multichain cost calculation}
\label{table2}
\centering 
\begin{tabular}{|c|c|c|c|c|}
\hline
 \textbf{Blockchain Network} & \textbf{\# of boats} & \textbf{Data point} &  \textbf{Total Cost in \$} \\
\hline
\hline
 EOS & 1000 &  10x365 & \$232 \\
\hline
\hline
 Stellar & 1000 & 10x365 &  \$197 \\
\hline
\hline
 Ethereum  & - & 2x365 & \$14 \\
\hline
\hline
 Grand total  &  &  & \$443 \\
\hline
\end{tabular}
\end{table}

In doing the computations, we assumed that each boat sends 10 significant data every day throughout one year and there are 1000 boats owned by the company.  Table \ref{table2} lists the costs associated with our proposed approach. It basically lists the costs relating to first level of blockchain (i.e., EOS and Stellar) for 1000 boats. For Ethereum since only the summary of data coming from EAS and Stellar is written, we have 2 per day only. The toal cost for our approach comes to \$443. 

\begin{table}[!ht]
\renewcommand{\arraystretch}{1.5}
\caption{Cost Comparison}
\label{table3}
\centering 
\begin{tabular}{|c|c|c|c|c|}
\hline
 \textbf{Aprroach} &  \textbf{Total Cost in \$} \\
\hline
\hline
 Multichain (EOS + Stellar + Ethereum) & \$443 \\
\hline
\hline
 Ethereum only (func. call) & \$13140 \\
\hline
\hline
 Ethereum only (new contract) & \$69350 \\
\hline
\end{tabular}
\end{table}

Table \ref{table3} lists the costs associated with other approaches compared to ours. As can be seen, the  cost  of Ethereum only  approach  is very expensive which is around \$70K. While it is highly secure and reliable, it will not be attractive for the boat company to deploy. The other Ethereum aproach with function calls turn out to be  much affordable around \$13K. This is because, the contract deployment cost is a one-time cost and the hashes are always written to this contract. Neverthless, this is still much expensive compared to our cost of \$443. The savings with our approach is significant and can be very attractive for the company to deploy. 


\section{Security Analysis}
In this section, we consider all the attacks mentioned in our Threat Model in Section IV and analyze how our proposed framework addresses these attacks. 

\textbf{Threat 1:}
In this scenario, the attacker tries to masquerade IoT device for pushing bogus IoT data into our multi-chain system. To do so, the attacker needs to derive the different private keys of IoT edge device that are used in EOS, Stellar and data center. We argue that even if the attacker may obtain one or more of these keys, the attack can be thwarted due to our MFI design. Any inconsistency between pushed data can be easily detected by the data center with a simple check. This means, the attacker needs to obtain all of the keys, which is very unlikely. 
Note that, we opt-out the stolen private key attack by assuming that HSM is deployed in IoT edge devices.

\textbf{Threat 2:}
Considering mentioned mempool features, the attacker may try to delete the transaction from the mempool. However, it is almost impossible because transaction pool is held by every node separately and the only way to delete these transactions is to remove them from all nodes in the network which means accomplishing a 51\% attack continuously for all blockchains in our multi-chain framework.

Another possibility is that the attacker can make too many bogus transactions with higher transaction fees to force nodes to remove the less paid transactions from their mempool. This attack has three main drawbacks. First, the attacker should invest huge amount of money to create enough bogus transactions to fill the mempools of all nodes for each blockchain. Second, this attack does not guarantee that only the related transactions (i.e., the ones whch hold the IoT hash values) will be removed from the mempool. Third, IoT device or data center can redo transactions if it is not confirmed within a reasonable time period.

\textbf{Threat 3:} In this attack scenario, the attacker may perform MitM attack between Blockchain peers and IoT edge devices. 
If the attack is successful, that means for both of the two mid-size Blockchain networks (i.e., EOS and Stellar), the attacker can block the transactions. However, at the end of the day when the data center is fetching the transactions to build the Merkle tree, the data center can easily figure out the problem and inform IoT edge device to push their IoT hash transactions again by using different EOS and/or Stellar nodes. 

\textbf{Threat 4:}
In our framework, Ethereum acts as an unbreakable seal to provide a long-term integrity ensuring mechanism for forensic investigations. This is due the fact that, Ethereum is a huge blockchain network which contains more than 10,000 full nodes. This makes Ethereum very secure against the 51\% attacks since the cost of such an attack is around \$400,000 per hour for now \cite{cost}. Thus, changing an old transaction, in other worlds, rolbacking will be worth \$400K $\times$ hours  depending on how old the transaction is. To change old transaction, the attacker must create a new and longer chain starting from the target block. Ethereum network has 215 TH/s hash rate, which is very high. To change old data in the Ethereum network requires calculating this difficulty from scratch for each succeeding blocks continuously.

\section{Conclusion}
\label{sec:conclusion}

In this paper, we proposed a forensics framework that consists of multiple blockchain networks in two layers. The purpose of the system is to verify authenticity and integrity of the data collected from various IoT devices in case of possible disputes. We collaboratively used multiple blockchains to create a more secure and tamper-resistance yet affordable system. For reducing the size of the data, we utilized hashes as well as Merkle tree to only store hash of hashes at the end of each day. 

We performed cost analysis with the actual prices obtained from three well-known blockchain networks and analyzed the security features of the design by considering possible attack scenarios. The results indicated that our framework reduces the costs significantly and makes it attractive to be used for companies. The system can be improved further by including additional low-cost blockchain platforms as they become available in the future to increase the resistance against possible attacks.

\bibliographystyle{IEEEtran}

\end{document}